\documentclass[lettersize,journal]{IEEEtran}
\usepackage{amsmath,amsfonts}
\usepackage{algorithmic}
\usepackage{algorithm}
\usepackage{array}
\usepackage[caption=false,font=normalsize,labelfont=sf,textfont=sf]{subfig}
\usepackage{textcomp}
\usepackage{stfloats}
\usepackage{url}
\usepackage{verbatim}
\usepackage{graphicx}
\usepackage{cite}
\begin{document}

\title{Interference Modulation: A Novel Technique for Low-Rate and Power Efficient Multiple Access}

\author{M. Yaser YAĞAN,~\IEEEmembership{Student Member,~IEEE,}
Ali E. PUSANE,~\IEEEmembership{Senior Member,~IEEE,}
Ali GÖRÇİN,~\IEEEmembership{Senior Member,~IEEE,}
İbrahim HÖKELEK,~\IEEEmembership{Member,~IEEE}
\thanks{M. Yaser YAĞAN is with Communications and Signal Processing Research (HİSAR) Laboratory, T{\"{U}}B{\.{I}}TAK B{\.{I}}LGEM, Kocaeli, Türkiye, and also with the Department of Electrical and Electronics Engineering, {Boğaziçi} University, İstanbul, Türkiye.  (e-mail:yaser.yagan@std.bogazici.gov.tr).}
\thanks{Ali E. PUSANE is with the Department of Electrical and Electronics Engineering, {Boğaziçi} University, İstanbul, Türkiye. (e-mail:ali.pusane@bogazici.edu.tr).}
\thanks{Ali GÖRÇİN is with Communications and Signal Processing Research (HİSAR) Laboratory, T{\"{U}}B{\.{I}}TAK B{\.{I}}LGEM, Kocaeli, Türkiye, and also with the Department of Electronics and Communication Engineering, Istanbul Technical University, {\.{I}}stanbul, Türkiye. (e-mail:aligorcin@itu.edu.tr).}
\thanks{İbrahim HÖKELEK is with Communications and Signal Processing Research (HİSAR) Laboratory, T{\"{U}}B{\.{I}}TAK B{\.{I}}LGEM, Kocaeli, Türkiye. (e-mail:ibrahim.hokelek@tubitak.gov.tr)}

}



\maketitle

\begin{abstract}
The majority of spatial signal processing techniques focus on increasing the total system capacity and providing high data rates for intended user(s). Unlike the existing studies, this paper introduces a novel interference modulation method that exploits the correlation between wireless channels to enable low-data-rate transmission towards additional users with a minimal power allocation. The proposed method changes the interference power at specific channels to modulate a low-rate on-off keying signal. This is achieved by appropriately setting the radiation pattern of front-end components of a transmitter, i.e., analog beamforming weights or metasurface configuration. The paper investigates theoretical performance limits and analyzes the efficiency in terms of sum rate. Bit error rate simulation results are closely matched with theoretical findings. The initial findings indicate that the proposed technique can be instrumental in providing reduced capability communication using minimal power consumption in 6G networks. 
\end{abstract}


\section{Introduction}
\IEEEPARstart{6}{G} use cases are expected to be diverse, including but not limited to the following innovative applications: holographic telepresence, emergency communication, e-health services, industrial robots/cobots, autonomous vehicles, and massive usage of reduced capability (RedCap) IoTs \cite{6G1}. 6G, in essence, needs to be sustainable in terms of green energy due to the increasing throughput demand and the massive number of end devices, resulting in high-power consumption in the radio network. These extremely challenging requirements for 6G networks have boosted research efforts in several directions. With the unprecedented acceleration in massive multiple-input-multiple-output (MIMO) technology and the fast evolution of metasurfaces, researchers have been exploiting the spatial domain and degrees of freedom (DoFs) provided by large antenna arrays \cite{MIMO1,MIMO2}. In this context, there have been numerous works for massive MIMO and beamforming to increase the capacity and improve the quality of service (QoS) with existing multiple access (MA) technologies \cite{MIMONew1,MIMONew2}. On the other hand, another group of studies has been dedicated to developing new MA techniques by utilizing spatial DoFs in multi-antenna systems, such as multi-user MIMO \cite{multiMIMO} and directional modulation \cite{DM,yaugan2024multi}. 

MIMO technology aims to exploit the orthogonality between channels to form parallel virtual links among pairs of transmitter and receiver antennas. Therefore, it is a technique to reduce the interference in principle, such that it can be utilized for transmitting multiple streams to a single user (diversity gain) or separate streams to multiple users (multiplexing gain). Besides the theoretical bounds where the number of data streams cannot exceed the number of dedicated radio frequency (RF) chains, practical limitations, in addition to the high hardware cost and energy consumption, prevent employing a high number of spatial layers in the MIMO technology. Furthermore, there has been a growing interest in full analog and hybrid beamforming architectures. Their ability to provide multiplexing gain and increase the throughput with lower hardware cost makes them preferable over conventional MIMO setups. To further increase the spectral efficiency of these architectures, new techniques such as index modulation (IM) \cite{IndexMod} and beam index modulation (BIM) \cite{BIMABF} were developed to convey additional information by means of the spatial domain. 

Opposite to MIMO, non-orthogonal multiple access (NOMA), which has been extensively studied, allows multi-stream transmission to users with comparable channel conditions by dividing another domain, such as power or code domains, and utilizing successive interference cancellation at the receivers. Consequently, it exploits interference to convey the information of users with different channel gains. While NOMA addresses some of the issues encountered in MIMO, it has its disadvantages, such as inefficient multiplexing gain and high transmitter/receiver complexity. 

In \cite{BIMABF}, BIM over a single RF chain with analog beamformers is proposed. BIM is developed as an improved spatial modulation technique that provides low cost and high energy efficiency at the expense of lower spectral efficiency. Although hardware complexity is reduced, specially designed receiver hardware is required to facilitate beam index detection. The work of \cite{APHBIM} has shown theoretically and numerically that combining the fully analog BIM with a precoding procedure to obtain a hybrid beamforming (HBF)-based BIM increases the spectral efficiency. 

In a more recent work \cite{mBIMFDA}, Qiu \textit{et al.} have developed a multi-beam transmission scheme that utilizes IM in frequency diverse arrays (FDA) to enhance physical layer security. This technique relies on the cooperation of legitimate users to recover confidential messages encoded in the transmitter's antenna indices. It also transmits artificial noise towards an unknown eavesdropper to ensure high security, hence it requires additional power for the artificial noise. 

With a focus on mmWave massive MIMO, a generalized beamspace modulation scheme was proposed in \cite{SMLRF}. This scheme enables multiplexing gain higher than the minimum number of RF chains utilized at the transmitter or the receiver. 

The authors in \cite{IRSmmBIM} proposed BIM with RIS to convey additional information for single-input-multiple-output (SIMO) systems at mmWave frequencies. Multi-RIS model, SNR optimization, and Rician channel model were investigated. The proposed scheme assumes direct access from the transmitter to the first RIS via a wired connection. It also requires a special receiver design, and hence, two detector structures were developed. 

While high throughput and low latency are key enablers for many 6G use cases, such as extended reality (XR) and autonomous systems, other applications like massive machine-type communications (mMTC) and RedCap devices place greater emphasis on low energy consumption and minimal complexity. As a result, there is a growing need to explore communication schemes optimized for efficiency, simplicity, and scalability—especially for scenarios where ultra-high performance is neither required nor practical.
\begin{figure}[]
\centerline{\includegraphics[width=.9\linewidth]{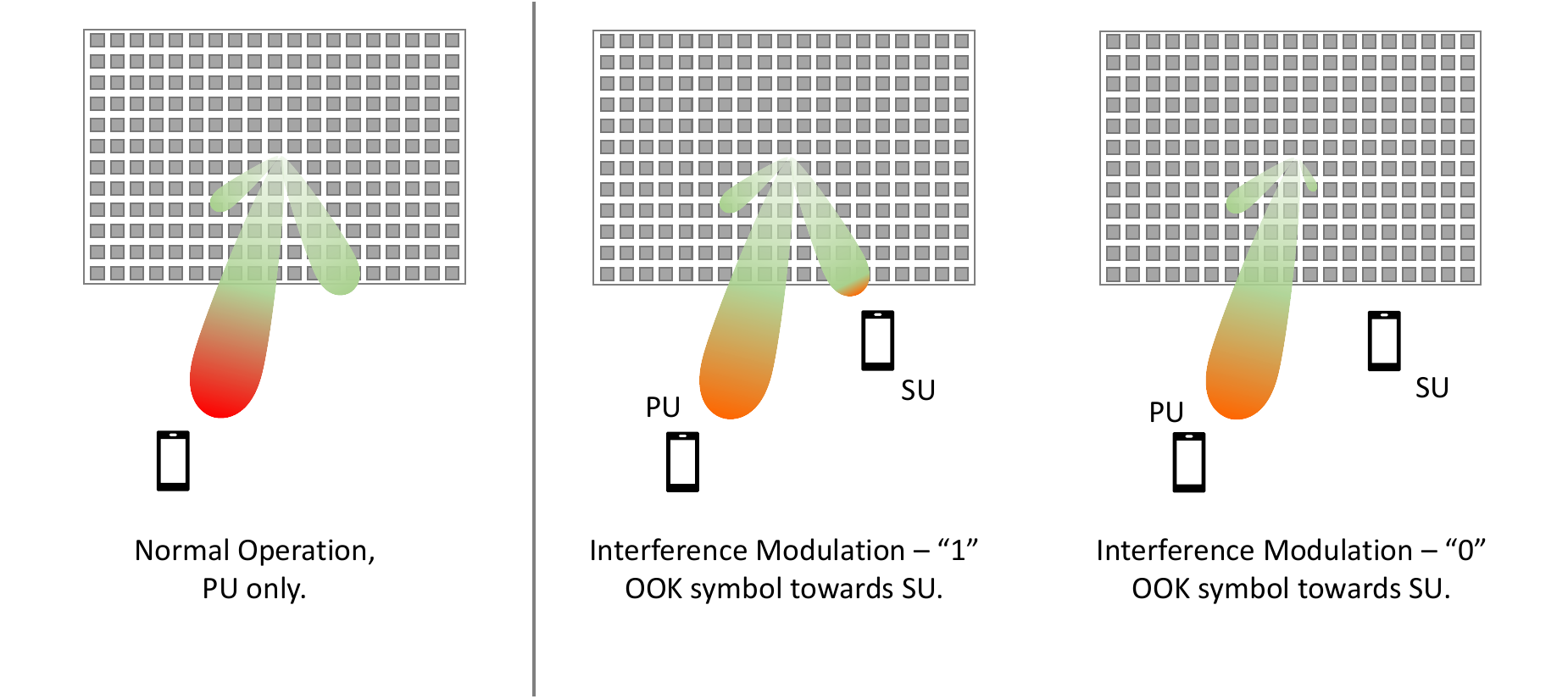}}
\caption{The proposed interference modulation concept.}
\label{fig:imconcept}
\end{figure}

In this paper, we propose a novel method for enabling low data rate transmission to a secondary user (SU) by modulating the interference from a high-bandwidth, orthogonal frequency division multiplexing (OFDM)-based transmission intended for a primary user (PU), as illustrated in Fig. \ref{fig:imconcept}. The key idea is to embed information in the interference pattern observed by the SU, whose channel is spatially correlated with that of the PU and has a similar gain. This is accomplished by dynamically adjusting the beamforming configuration or precoding matrix at the transmitter, thereby altering the interference level at the SU in a controlled manner. Theoretical analysis supported by simulations shows that the proposed scheme can increase the sum rate under certain channel conditions. Notably, this approach requires no additional hardware, making it highly attractive for scenarios where energy and complexity constraints are critical, such as in RedCap communications.

The rest of this paper is organized as follows: Section~\ref{sec:system_model} introduces the system model where the proposed interference modulation scheme can be utilized, and Section~\ref{sec:interference_mod} describes how the interference modulation can be implemented, including parameter calculation and bit error analysis. Numerical results with efficiency evaluation are given in Section~\ref{sec:results}, and the paper is concluded in Section~\ref{sec:conclusion}.
\section{System Model}
\label{sec:system_model}

\begin{figure}[]
\centerline{\includegraphics[width=1\linewidth]{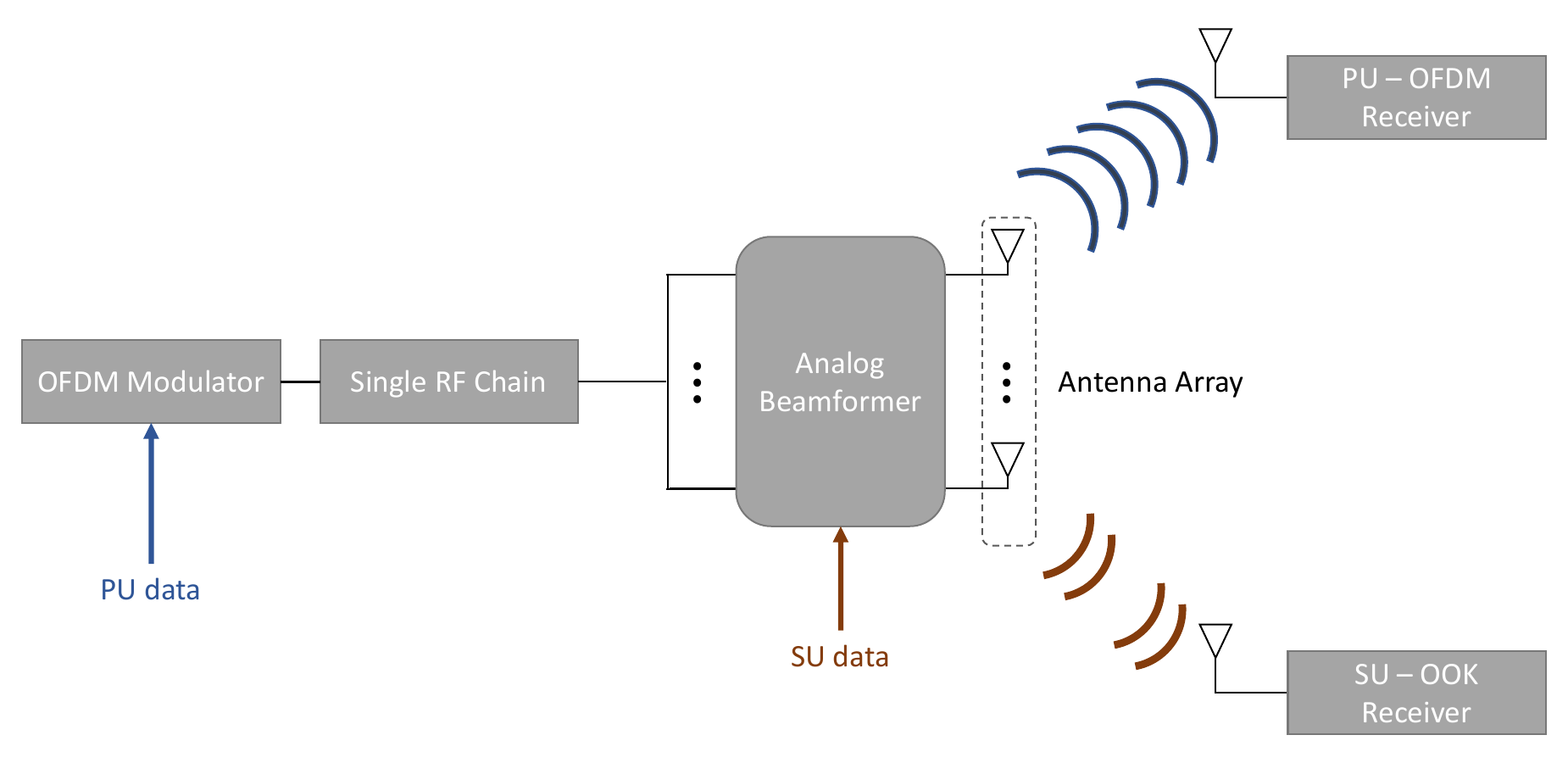}}
\caption{The proposed modulator model.}
\label{fig:imarchitecture}
\end{figure}

The paper proposes and investigates interference modulation for the system architecture shown in Fig. \ref{fig:imarchitecture}. The model assumes a single-RF chain transmitter with $K$ antennas serving a single PU and will implement interference modulation to allocate a portion of the PU's power for a SU.  
The single-antenna PU is assumed to be receiving an OFDM waveform with a known power level, the received signal can be expressed as
\begin{equation}
    y_{_{PU}} = \boldsymbol{h_{PU}}^T ~\boldsymbol{\omega}~ s + n_{_{PU}},
\end{equation}
with $\boldsymbol{h_{PU}}$ is the channel vector between the $K$ transmitter antennas and the receiver, $\boldsymbol{\omega}$ is the beamforming weight vector, $s$ is the transmitted OFDM signal, and $n_{_{PU}}$ represents the additive white Gaussian noise with zero-mean normal distribution and $\sigma_n^2$ variance $n_{_{PU}}\sim \mathcal{N}(0,\sigma_n)$.
When a SU is present, the same transmitted signal will be received as
\begin{equation}
    y_{_{SU}} = \boldsymbol{h_{SU}}^T ~\boldsymbol{\omega}~ s + n_{_{SU}}, 
    \label{eq:interference}
\end{equation}
with $\boldsymbol{h_{SU}}$ denoting the channel between the transmitter and the SU, and $n_{_{SU}}\sim \mathcal{N}(0,\sigma_n)$.

For interference modulation, two different beamforming weight vectors ($\boldsymbol{\omega_1},\boldsymbol{\omega_0}$) will be dynamically applied at the transmitter. 
The PU will experience a relatively low drop in the received power level. This power will be repeatedly focused towards the SU (with $\boldsymbol{\omega_1}$) or scattered (with $\boldsymbol{\omega_0}$) to deliver a modulated on-off keying (OOK) one and OOK zero symbols, respectively, with a lower bandwidth. The PU signal delivered to the SU ($y_{_{SU}}$) will be strong enough to maintain proper operation of an energy detector-based receiver, and relatively weak to not compromise the PU's information carried on the waveform. 
Since this model realizes an OOK transmission towards the SU, the spectral efficiency is bounded by 1 bit/s/Hz. Furthermore, the allocated power for this transmission should be calculated precisely such that the loss in PUs capacity should not be higher than the aforementioned spectral efficiency.

\section{Interference Modulation}
\label{sec:interference_mod}
In normal operation, where the transmitter is focusing the signal in the direction of PU, a portion of the signal is received by the SU as interference, as given in (\ref{eq:interference}). The level of this interference depends on the correlation between the channels $\boldsymbol{h_{PU}}$ and $\boldsymbol{h_{SU}}$ and their gain difference. 
Interference modulation configurations ($\boldsymbol{\omega_1},\boldsymbol{\omega_0}$) should be implemented without affecting the PU received signal, or they can reduce its power slightly, but not with a notable change when the configuration changes, and this is the first problem to investigate.
\subsection{Power Allocation and Efficiency}
To maintain a desired error probability for the received OOK symbols at the SU, a power coefficient $\alpha$ should be defined and allocated for the SU channel such that $\left|\boldsymbol{h_{SU}}^T \boldsymbol{\omega_1}\right| = \sqrt{\alpha}$ and $\left|\boldsymbol{h_{PU}}^T \boldsymbol{\omega_1}\right| = \sqrt{1-\alpha}$ when a logical one OOK symbol is transmitted. For the PU not to be affected, $\left|\boldsymbol{h_{SU}}^T \boldsymbol{\omega_0}\right| = 0$ and $\left|\boldsymbol{h_{PU}}^T \boldsymbol{\omega_0}\right| = \sqrt{1-\alpha}$ should also be satisfied when logical zero is transmitted. The channel gains of $\boldsymbol{h_{PU}}$ and $\boldsymbol{h_{SU}}$ can be ignored in these calculations as they can be implicitly taken into account in the receivers' SNR. Thus, $\boldsymbol{h_{PU}}$ and $\boldsymbol{h_{SU}}$ can be assumed to have unit norm.

If the number of transmitter antennas is greater than 2, then satisfying one of the mentioned cases is equivalent to solving an underdetermined equation system. A minimum norm solution can be found using the Moore–Penrose pseudoinverse method \cite{moon2000mathematical}. However, for power constraints, the average norm of $\boldsymbol{\omega_1}$ and $\boldsymbol{\omega_0}$ must not exceed 1. The squared norm of the solution to any of the two mentioned equation systems can be expressed as
\begin{equation}
    {\left|\boldsymbol{\omega_k}\right|}^2 = \boldsymbol{\omega_k}^H \boldsymbol{\omega_k} = \boldsymbol{b}^H {{\left(\boldsymbol{A}^H\boldsymbol{A}\right)}^{-1}}^{H}\boldsymbol{A}^H \boldsymbol{A} {\left(\boldsymbol{A}^H\boldsymbol{A}\right)}^{-1}\boldsymbol{b},
    \label{eq:norm1}
\end{equation}
where $\boldsymbol{b} = {[b_{SU,k} ~~~~b_{PU} ]}^T$, $\boldsymbol{A} = {[\boldsymbol{h_{SU}} ~~~~ \boldsymbol{h_{PU}} ]}$, and $k\in\{0,1\}$. $b_{PU}$ is a complex number that satisfies $|b_{PU}| = \sqrt{1-\alpha}$ and $b_{SU,k}$ satisfies $|b_{SU,1}| = \sqrt{\alpha}$ and $ b_{SU,0} = 0$ according to the previous explanation. Eq. \ref{eq:norm1} can be simplified to
\begin{multline}
    {\left|\boldsymbol{\omega_k}\right|}^2 = \boldsymbol{b}^H {{\left(\boldsymbol{A}^H\boldsymbol{A}\right)}^{-1}}^{H}\boldsymbol{b} \\
    =
    \begin{bmatrix} b_{SU,k}^* & b_{PU}^*\end{bmatrix} \frac{1}{1-{|\rho|}^2} \begin{pmatrix}
        1 & -\rho^* \\ -\rho & 1
    \end{pmatrix}
    \begin{bmatrix}
        b_{SU,k} \\ b_{PU}
    \end{bmatrix} \\
    = \frac{|b_{PU}|^2 + |b_{SU,k}|^2 -2\mathfrak{R}\{\rho b_{SU,k}^{~} b_{PU}^*\}}{1-{|\rho|}^2},
\end{multline}
where $\mathfrak{R}\left\{\cdot\right\}$ denotes the real part of the complex-valued argument, and $\rho = <\boldsymbol{h_{SU}},\boldsymbol{h_{PU}}>$ is the inner product of the unit-norm channel vectors representing their correlation. While the last term in the numerator disappears for $k=0 \to b_{SU,0} = 0$, this norm can be minimized for $k=1$ by aligning the phases of the complex values $(\rho b_{SU,k}^{~} b_{PU}^*)$. Let the channel correlation term be $\rho=|\rho| e^{j\theta}$, then $b_{SU,1}$ and $b_{PU}$ can be set to $(\sqrt{\alpha}e^{-j\theta})$ and $(\sqrt{1-\alpha})$, respectively. As a result, the interference modulation weights and their norms are found in terms of $\alpha$ and $\rho$ with $\arg(\cdot)$ being the phase of the complex number as
\begin{equation}
    \begin{matrix}
        \boldsymbol{\omega_0} = \boldsymbol{A}{\left(\boldsymbol{A}^H\boldsymbol{A}\right)}^{-1} \begin{bmatrix}
            0 \\ \sqrt{1-\alpha}
        \end{bmatrix},
        \\ \\  {\left|\boldsymbol{\omega_0}\right|}^2 = \frac{1-\alpha }{1-{|\rho|}^2}, \\ \\ 
        \boldsymbol{\omega_1} = \boldsymbol{A}{\left(\boldsymbol{A}^H\boldsymbol{A}\right)}^{-1} \begin{bmatrix}
            \sqrt{\alpha} ~e^{j\arg(\rho)} \\ \sqrt{1-\alpha}
        \end{bmatrix},
        \\ \\  {\left|\boldsymbol{\omega_1}\right|}^2 = \frac{1-\sqrt{\alpha}\sqrt{1-\alpha}|\rho| }{1-{|\rho|}^2}.
    \end{matrix}
    \label{eq:norm2}
\end{equation}

\begin{figure}[]
\centerline{\includegraphics[width=.9\linewidth]{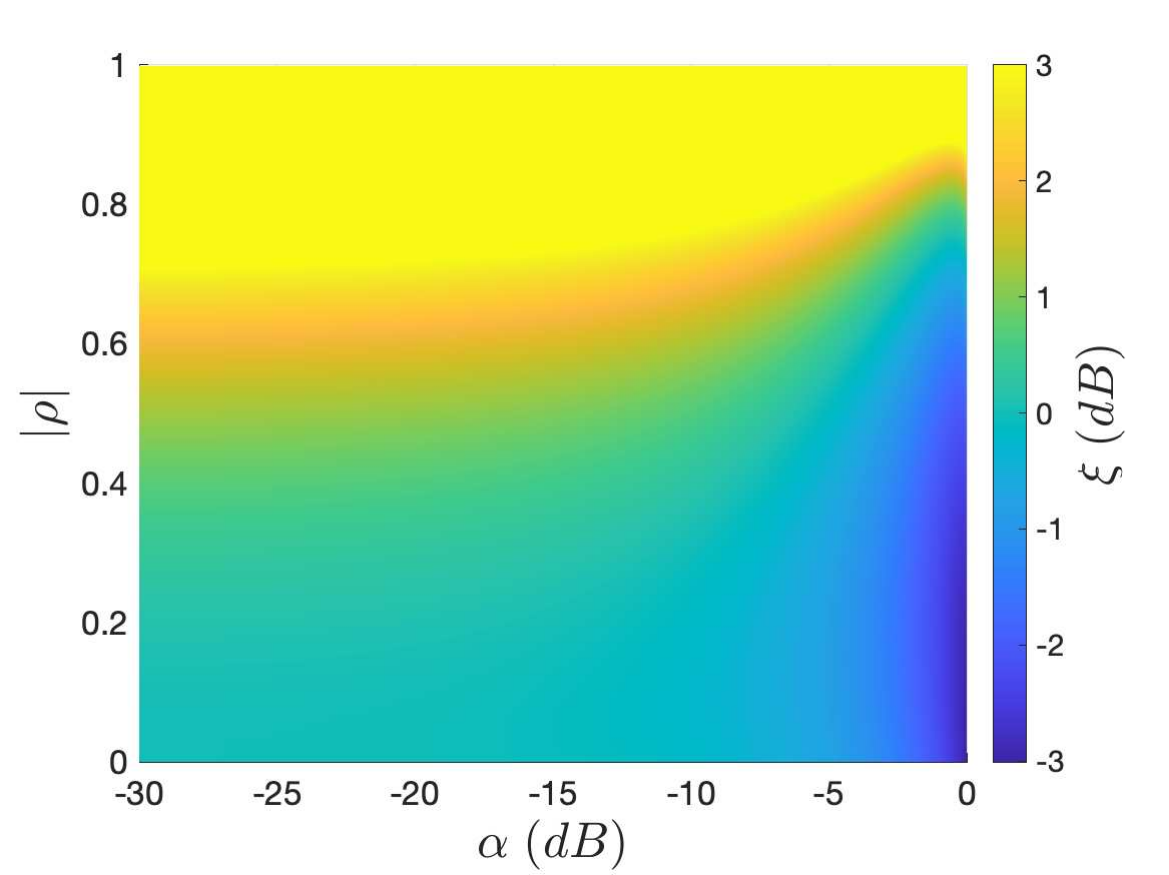}}
\caption{The power loss in the proposed model depending on the channels' correlation and SU power coefficient.}
\label{fig:efficiency}
\end{figure}

For power constraints, the mean beamforming weights norm should not exceed a predefined value. Assuming unit norm for normal operation weights, $\boldsymbol{\omega_0}$ and $\boldsymbol{\omega_1}$ will be normalized with respect to their average $\xi$, which is obtained as
\begin{equation}
    \xi = \frac{2-\alpha-\sqrt{\alpha}\sqrt{1-\alpha}|\rho|}{2-2|\rho|^2}.
\end{equation}
If $\xi$ is greater than one, this means that PU and SU after normalization will receive power less than $1-\alpha$ and $\alpha$, respectively, indicating some power loss to satisfy the channel conditions. On the other hand, having $\xi <1$ will result in a higher received power, meaning a gain in the modulation. Furthermore, $\xi$ can be used as an efficiency metric, and (\ref{eq:norm2}) shows that this metric is calculated by means of normalized channel correlations and the SU power allocation coefficient $\alpha$, and is independent of the channels themselves. A graphical representation of $\xi$ is shown in Fig. \ref{fig:efficiency}. This is the amount of consumed power to mathematically satisfy the required interference level at the SU characterized by $\alpha$. Thus, a method to define $\alpha$ such that a clear OOK signal is observed at the SU is required. This method will be developed upon the calculation of the bit error probability of the SU.

\subsection{SU's Bit Error Probability}
To analyze the bit error probability, the probability density functions (PDFs) of received samples for OOK symbols will be calculated.
An OFDM time sample follows a zero-mean normal distribution as the summation of a large number of subcarriers converges to a normal distribution according to the central limit theorem. Thus, the PDF for its real and imaginary parts is 
\begin{equation}
    f_S(s) = \frac{1}{\sqrt{\pi\sigma^2}}e^{\frac{x^2}{\sigma^2}},
\end{equation}
where $\sigma^2=1/M$ and $M$ is the number of the subcarriers. With zero-mean additive white Gaussian noise, the sample will be observed at the receiver as the summation of two Gaussian random variables with variances $\sigma_r^2$ and $\sigma_n^2$, which also follows the normal distribution. $\sigma_r^2$ presents the received OFDM signal power, which will be calculated as
\begin{equation}
    \sigma_r^2 = \sigma^2\frac{\alpha {|\boldsymbol{\omega_1}|}^2}{\xi} g^2,
\end{equation}
where $g$ represents the channel gain ratio between SU and PU channels, and the fractional term describes the interference modulation gain. The instantaneous power of this waveform, which is calculated as $p = xx^*$ with $x^*$ being the complex conjugate of the noisy time samples $x$, has an exponential distribution as 
\begin{equation}
    f_P(p) = \frac{e^{-\frac{p}{(\sigma_r^2 + \sigma_n^2)}}}{(\sigma_r^2 + \sigma_n^2)}.
\end{equation}

\begin{figure}[]
\centerline{\includegraphics[width=.8\linewidth]{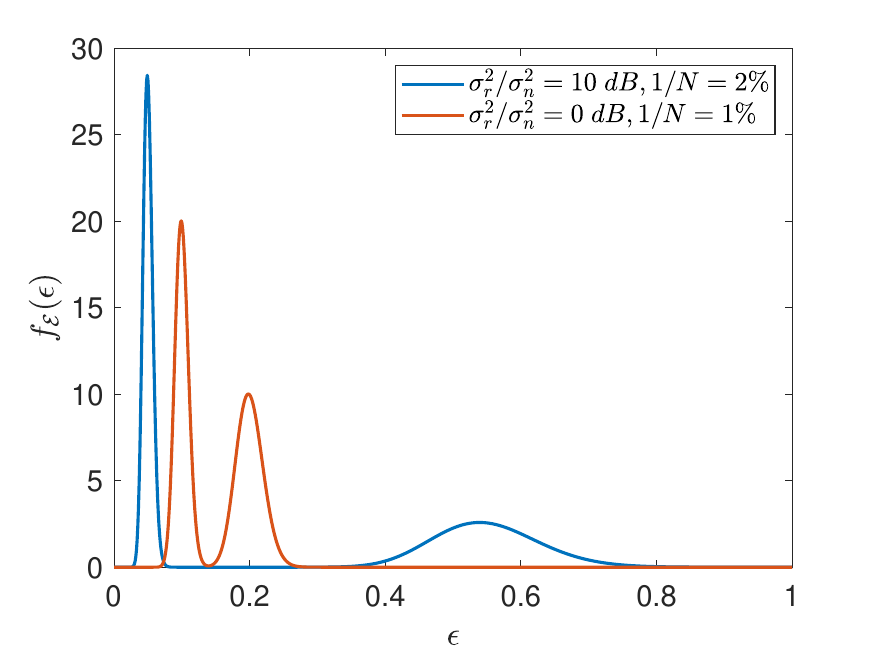}}
\caption{Example PDFs of the received signal energy for OOK.}
\label{fig:expdfs}
\end{figure}

The total energy observed at the SU for a period of $N$ samples is 
\begin{equation}
    \epsilon = \sum_N p_i,
\end{equation}
and it has a Gamma distribution with a shape parameter $N$ and scale parameter $(\sigma_r^2 + \sigma_n^2)$ as
\begin{equation}
    f_\mathcal{E}(\epsilon) = \frac{1}{\Gamma(N){(\sigma_r^2 + \sigma_n^2)}^N} \epsilon^{N-1}e^{-\epsilon/(\sigma_r^2 + \sigma_n^2)}.
\end{equation}
On the other hand, when no OFDM signal is received, only the noise energy will be observed. Assuming an equal probability of zeros and ones, the total energy probability distribution function will be 
\begin{multline}
    f_\mathcal{E}(\epsilon) = 0.5\frac{1}{\Gamma(N){(\sigma_r^2 + \sigma_n^2)}^N} \epsilon^{N-1}e^{-\epsilon/(\sigma_r^2 + \sigma_n^2)} + \\  0.5 \frac{1}{\Gamma(N){\sigma_n^2}^N} \epsilon^{N-1}e^{-\epsilon/\sigma_n^2}.
    \label{eq:energy}
\end{multline}
Eq. (\ref{eq:energy}) shows that the shapes of the received signal energy probability distributions depend on the number of observed samples $(N)$, which defines the ratio between SU communication rate and the PU communication bandwidth, and the signal-to-noise ratio of the OFDM signal $(\sigma_r^2/\sigma_n^2)$. Examples are shown in Fig. \ref{fig:expdfs}. 

To detect the transmitted OOK symbols, a threshold $\delta$ is calculated to minimize the detection error, where the error probability is 
\begin{multline}
    P_e = 0.5\int_\delta^\infty \frac{1}{\Gamma(N){\sigma_n^2}^N} \epsilon^{N-1}e^{-\epsilon/\sigma_n^2} d\epsilon + \\ 0.5\int_{0}^\delta \frac{1}{\Gamma(N){(\sigma_r^2 + \sigma_n^2)}^N} \epsilon^{N-1}e^{-\epsilon/(\sigma_r^2 + \sigma_n^2)} d\epsilon.
\end{multline}
    
Taking the derivative of the error probability with respect to $\delta$ and equating it to zero, the error minimizing threshold value is calculated as
\begin{equation}
    \delta^* = \frac{{N\left(\ln{\left(\sigma_r^2 + \sigma_n^2\right)}-\ln{\left(\sigma_n^2\right)}\right)}}{1/(\sigma_r^2 + \sigma_n^2)-1/\sigma_n^2},
\end{equation}
and the corresponding error probability is 
\begin{multline}
    P_e = 0.5 \left(1 - \frac{1}{\Gamma(N)}\gamma\left(N,\delta^*/\sigma_n^2\right) + \right. \\ \left. \frac{1}{\Gamma(N)}\gamma\left(N,\delta^*/(\sigma_r^2 +\sigma_n^2)\right)\right),
    \label{eq:Pe}
\end{multline}
where $\gamma()$ is the lower incomplete Gamma function defined as $\gamma(s,x) = \int_0^xt^{s-1}e^{-t}dt$. 

In these calculations, it is assumed that the SU's bandwidth and sampling rate are identical to those of the PU. However, they are not required to be the same. If the SU's bandwidth is lower than the PU's bandwidth, then the SU will observe a part of the transmitted signal's power. Typically, this should not change the SNR, since the received signal power and the noise power will be reduced by the same ratio.

\section{Numerical Results}
\label{sec:results}
\begin{figure}[]
\centerline{\includegraphics[width=.8\linewidth]{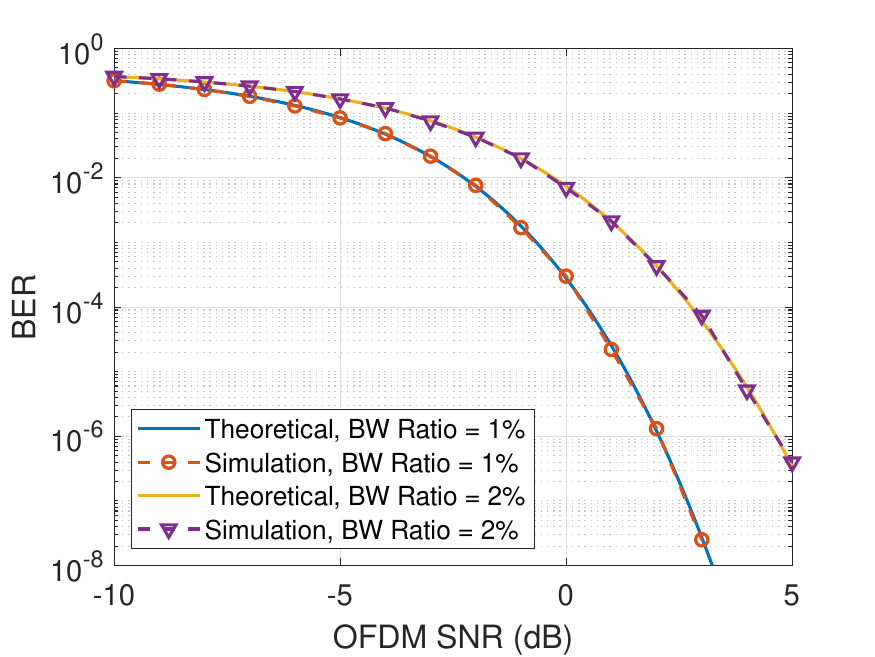}}
\caption{Simulated BER and theoretical error probability.}
\label{fig:ber}
\end{figure}

The bit error rate (BER) simulation for the proposed interference modulation scheme has been conducted for various values of $(\sigma_r^2/\sigma_n^2)$ and for two different values of bandwidth ratio. The simulation results are consistent with the theoretical error probability, as shown in Fig. \ref{fig:ber}. 

\begin{figure}[]
\centerline{\includegraphics[width=.8\linewidth]{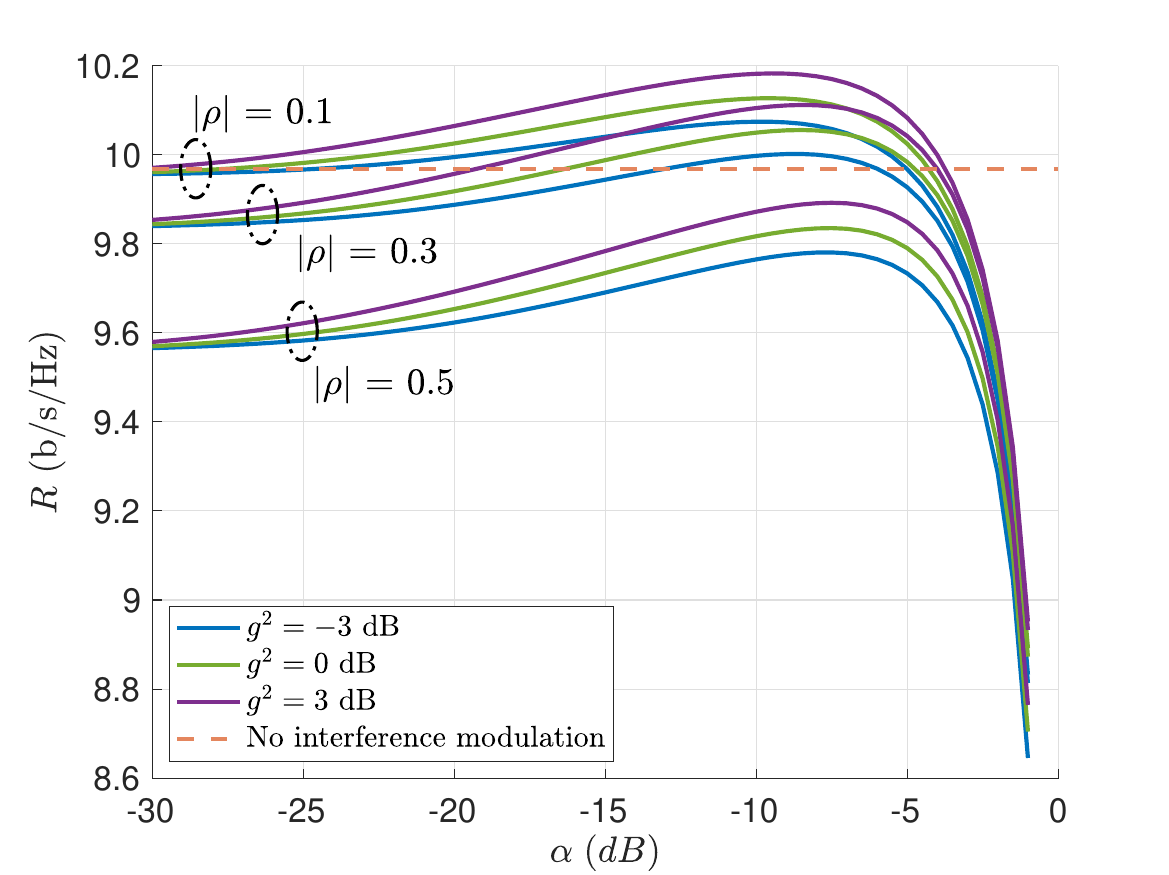}}
\caption{Sum rate for the proposed modulator with $\gamma = 30$ dB.}
\label{fig:sum_rate}
\end{figure}

Having the bit error probability calculated, the transmitter can determine the bandwidth ratio (in terms of $N$) and the allocated power (in terms of $\alpha$) to provide a low-error transmission. Then, according to the chosen $\alpha$ value and resulting $\xi$ value, the loss in PU capacity can be calculated. As a result, denoting the PU's SNR under normal operation (no interference modulation) as $\gamma$, the sum rate is calculated as 
\begin{equation}
    R = \log_2\left(1+\frac{\gamma}{\xi}(1-\alpha)\right) + 1/N_\alpha.
\end{equation}
$N_\alpha$ denotes the highest bandwidth ratio that achieves low error probability ($P_e < 10^{-5}$) for a given $\alpha$. It is numerically calculated by evaluating (\ref{eq:Pe}) for various values of $N$ until finding $N_\alpha$. Examples of resulting sum rate for different SU/PU channel gain ratios and correlations are shown in Fig. \ref{fig:sum_rate}. $\gamma$ is taken as $30$ dB, where the PU capacity in normal operation is $9.97$ b/s/Hz. As seen in the figure, the most dominating element in the efficiency of the modulator is the channel correlation. Higher correlation results in higher power loss, especially for small $\alpha$ coefficients, since the correlation results naturally in high interference. Additionally, since small $\alpha$ values mean lower OFDM SNR at the SU, the achievable error-free transmission bandwidth is minimized, and thus the increase in the sum rate is minor. In each curve, the sum rate increases until a certain $\alpha$ value, where the loss in the PU capacity is much lower than the gain in the SU rate. Then it starts to decrease rapidly due to the increasing loss in the PU capacity. In summary, interference modulation increases the spectral efficiency for channels with similar gains and low correlation. 
\section{Conclusion}
This paper proposes interference modulation, a novel technique that enables simultaneous transmission towards two different users in a single-RF chain multi-antenna system. The transmitter serves a primary user with a high-bandwidth OFDM waveform and modulates its interference to a secondary user to deliver low-rate OOK symbols. Theoretical analysis and simulation results show that this scheme can increase the sum rate under certain channel conditions, paving the way to a new approach for energy-efficient RedCap transmission with minimal power allocation. The proposed model can be applied to metasurface structures such as reconfigurable intelligent surfaces and reconfigurable holographic surfaces. Experimental verification will be conducted and reported in a future work along with the implementation in MIMO with hybrid beamforming architectures will also be investigated.   
\label{sec:conclusion}

\section{Acknowledgment}
This work has received funding from the EcoMobility project. EcoMobility has been accepted for funding within the Key Digital Technologies Joint Undertaking (KDT JU) which is rebranded as CHIPS, a public-private partnership in collaboration with the HORIZON Framework Programme and the national Authorities under grant agreement number 101112306.

\bibliographystyle{IEEEtran}

\bibliography{refs}
\end{document}